# A pathway to optimize the properties of magnetocaloric $Mn_xFe_{2-x}(P_{1-y}Ge_y)$ for magnetic refrigeration

D. M. Liu[a,#], Z.L.Zhang[a], S. L. Zhou[b], Q. Z. Huang[c], X. J.Deng[a], M. Yue[b], C. X. Liu[a], F. X.Hu[d], G. H. Rao[d], B. G. Shen[d], J. X. Zhang[b], and J. W. Lynn[c,*]

[a] Institute of Microstructure and Property of Advanced Materials, Beijing University of Technology, 100 Pingleyuan, Chaoyang District, Beijing 100124, China

[b] College of Materials Science and Engineering, Beijing University of Technology, 100 Pingleyuan, Chaoyang District, Beijing 100124, China

[c] NIST Center for Neutron Research, National Institute of Standards and Technology, Gaithersburg, Maryland 20899 (USA)

[d] Beijing National Laboratory for Condensed Matter Physics, Institute of Physics, Chinese Academy of Sciences, Beijing 100190, China

## Abstract

Magnetocaloric materials can be useful in magnetic refrigeration applications, but to be practical the magneto-refrigerant needs to have a very large magnetocaloric effect (MCE) near room temperature for modest applied fields (<2 Tesla) with small hysteresis and magnetostriction, and should have a complete magnetic transition, be inexpensive, and environmentally friendly. One system that may fulfill these requirements is $Mn_xFe_{2-x}P_{1-y}Ge_y$, where a combined first-order structural and magnetic transition occurs between the high temperature paramagnetic and low temperature ferromagnetic phase. We have used neutron diffraction, differential scanning calorimetry, and magnetization measurements to study the effects of Mn and Ge location in the structure on the ordered magnetic moment, MCE, and hysteresis for a series of compositions of the system near optimal doping. The diffraction results indicate that the Mn ions located on the 3$f$ site enhance the desirable properties, while those located on the 3g sites are detrimental. The entropy changes measured directly by calorimetry can exceed 40 J/kg·K. The phase fraction that transforms, hysteresis of the transition, and entropy change can be controlled by both the compositional homogeneity and the particle size, and an annealing procedure has been developed that substantially improves the performance of all three properties of the material. On the basis of these results we have identified a pathway to optimize the MCE properties of this system for magnetic refrigeration applications.

___________________________________________________________________________
*Corresponding author. Tel.+1 3019756246. FAX: +1 301 921-9847
E-mail address: jeffrey.lynn@nist.gov (J.W. Lynn).

#Tel.: +86 1067391761. Fax: +86 1067396611
E-mail address: dmliu@bjut.edu.cn (D.M. Liu).

# 1. Introduction

Magnetic refrigeration based on the magnetocaloric effect (MCE) has attracted recent interest as a potential replacement for the classical vapor compression systems in use today. For a transition that is purely magnetic in origin, which is typically second order (continuous) in nature, the MCE is insufficient. To increase the entropy change at the phase transition, systems with concomitant magnetic and structural transitions, which are first order (discontinuous), can greatly increase the MCE, but at the cost of hysteresis and transitions that are phase-incomplete. To be considered for practical application in magnetocaloric refrigeration the first-order transition must have: (i) a large entropy change $\Delta S$ associated with the transition from a disordered to the ordered magnetic structure; (ii) the transition must occur near room temperature, (iii) must have a small hysteresis $\Delta T_{hys}$; (iv) a low applied magnetic field $\Delta B$ to induce the transition; iv) a small temperature/applied magnetic field range of co-existence of the ordered and disordered phases; (vi) and have a complete magnetic transition, i.e. there should be no, or very small, residual untransformed material when the process is completed.

The choice of the materials studied so far, however, has been made mainly on the basis of the magnitude of the entropy change. Some materials with transitions near room temperature have been $Gd_5Si_2Ge_2$ [1], $MnFeP_{0.45}As_{0.55}$ [2] and $LaFe_{13-x}Si_x$ [3]. In all these cases the detailed structures of the components were not reported and, consequently, information was not provided about the temperature/applied field range of coexistence of the two phases and about the completeness of the transition. In addition, in these early works only nominal compositions were reported and no structural details such as refined compositions were given, making it impossible to establish correlations between the above factors and to obtain precise indications about the most favorable compositions and preparation conditions of these materials.

The $Mn_xFe_{2-x}(P_{1-y}Ge_y)$ material is one such system with a combined structural and magnetic transition from the paramagnetic (PM) to ferromagnetic (FM) state. It has a high MCE and a chemical composition that avoids the use of expensive elements such as Gd in $Gd_5(Si_xGe_{1-x})_4$ or toxic ones such as As in $MnFe(P_{1-x}As_x)$ [4-6]. Trung, *et al.* [7] reported a comprehensive study for a variety of compositions in the $Mn_xFe_{2-x}P_{1-y}Ge_y$ system and found that the thermal hysteresis can be tunable. They also showed that the value of the Curie temperature ($T_C$) increases when the Ge content increases and decreases when the Mn content increases. Leitão, *et al.* reported [8] that the $T_c$ of the Fe rich side of the $(Mn,Fe)_2(P,Ge)$ system is easy to tune with careful manipulation of Fe and Ge content, but the ferro–paramagnetic transition sharpness decreases in the compounds [8]. In our previous work [9-11] the crystal and magnetic structures for the compound with composition $Mn_{1.1}Fe_{0.9}(P_{0.8}Ge_{0.2})$ and $Mn_{1.1}Fe_{0.9}(P_{0.76}Ge_{0.24})$ were determined by neutron powder diffraction (NPD). These compounds have the $Fe_2P$-type structure (space group $P\bar{6}2m$) and undergo a combined first-order structural and magnetic transition from a paramagnetic (PM) to a ferromagnetic (FM) phase. In those experiments we found that the two phases coexist in an interval of applied magnetic-field/temperature, and that the range of coexistence depends on the composition. In general, the transition does not go to completion so that there is a remnant untransformed fraction of material. This behavior was attributed to preparation conditions and inhomogeneities of composition in the samples. However, no detailed results were obtained to systematically investigate the relationship between properties and methods of preparation of the samples. In this paper we investigate this problem in $Mn_{2-x}Fe_xP_{1-y}Ge_y$ for a number of compositions, with the goal of elucidating in detail the effects of composition, method of preparation, and crystal structure on the physical properties, so that the correlation between the composition and physical properties can be established, and an optimized composition identified.

## 2. Experimental

Mn$_{2-x}$Fe$_x$P$_{1-y}$Ge$_y$ (x=0.8, 0.9; y=0.2, 0.22, 0.24) powders were prepared by ball milling. The polycrystalline samples were obtained by subsequent spark plasma sintering method as described in previous work [12]. Detailed temperature and magnetic field neutron diffraction measurements were carried out on the high-intensity BT9 triple axis spectrometer at NIST Center for Neutron Research (NCNR) using pyrolytic graphite monochromator and filter, with a wavelength of 2.359 Å. Neutron powder diffraction data were also collected on the high resolution powder neutron diffractometer (BT1), with monochromatic neutrons of wavelength 1.5403 Å produced by a Cu(311) monochromator. Magnetic field measurements were carried out with a vertical field 7 T superconducting magnet.

The temperature and field dependences of the magnetization were measured with a Quantum Design superconducting quantum interference (SQUID) instrument. The temperature steps were 1 K apart closer to the transition and 2 K further away. A Netzsch differential scanning calorimetry 204 F1 was used for calorimetric measurements. The sample of about 30 mg was investigated by heating at the rate of 1 K/min.

X-ray temperature-dependent diffraction measurements were carried out with a Bruker D8 Advance equipment. Scanning Electron Micrsopy (SEM) using a FEI Quanta 250, and Electron Backscatter Diffraction (EBSD) using a EDAX/TSL system, were used to investigate the microstructure of the components. Energy-dispersive X-ray spectroscopy (EDAX) and scanning transmission synchrotron radiation X-ray microscope experiments were used for elemental distribution analysis.

## 3. Results and discussion
### 3.1. Crystal and magnetic structures

The crystal and the magnetic structures of Mn$_x$Fe$_{2-x}$P$_{1-y}$Ge$_y$ are shown in Fig. 1. Mn$_x$Fe$_{2-x}$P$_{1-y}$Ge$_y$ adopts the hexagonal Fe$_2$P-type structure with partial substitution of Mn for Fe and Ge for P. The crystal structures of the Mn$_{2-x}$Fe$_x$P$_{1-y}$Ge$_y$ samples were refined in the hexagonal space group $P\bar{6}2m$ with the GSAS program [13] using neutron powder diffraction data collected at 295 K. Nominal and refined compositions are given in Table 1. In the alloys the refined Mn compositions are slightly less than the nominal ones while the Fe content is slightly higher. In the structure of Mn$_x$Fe$_{2-x}$P$_{1-y}$Ge$_y$, for x<1, most or all the Mn atoms are located in the 3g sites. When x>1, most or all the Mn atoms in excess of 1 occupy the 3f sites and will be indicated as % of Mn in 3f with the symbol n(Mn)%3f, where n is the site occupancy factor of Mn. It is found that n(Mn)%3f has a corresponding relation with some important magnetocaloric properties as discussed below. The substitution of Mn for Fe atoms in the 3f site (n%Mn-3f) is also provided in Table 1. Note that the n%Mn-3f depends on the Mn/Fe ratio. Substitution of Ge for P leads to a random distribution of Ge in each of the 1b and 2c sites, with the Ge preferentially locating in the 2c sites.

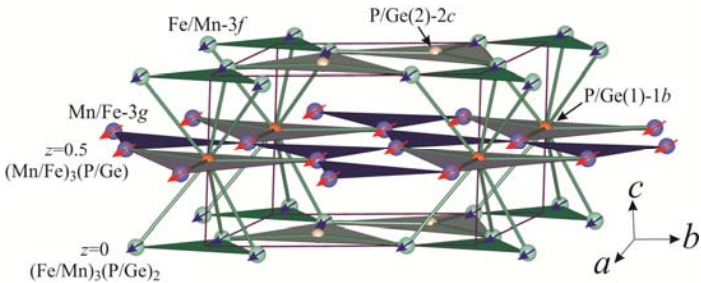

Fig. 1. (color online) Crystal and magnetic structure of Mn$_x$Fe$_{2-x}$P$_{1-y}$Ge$_y$.

**Table 1.** Nominal and refined compositions, substitution of Mn for Fe atoms in the 3f site (n%Mn-3f) of the samples. Refined compositions were obtained by NPD on the sintered bulk samples.

| Sample | Nominal composition | Refined composition | n%Mn at 3f site at 295k |
|---|---|---|---|
| LZ091- 1# | $Mn_{1.2}Fe_{0.8}P_{0.76}Ge_{0.24}$ | $Mn_{1.17}Fe_{0.83}(P_{0.74}Ge_{0.26})$ | 17.03 |
| LZ092 | $Mn_{1.2}Fe_{0.8}P_{0.76}Ge_{0.24}$ | $Mn_{1.17}Fe_{0.83}(P_{0.74}Ge_{0.26})$ | 17.4 |
| LZ094 | $Mn_{1.2}Fe_{0.8}P_{0.76}Ge_{0.24}$ | $Mn_{1.16}Fe_{0.84}(P_{0.76}Ge_{0.24})$ | 16.4 |
| LL083 | $Mn_{1.1}Fe_{0.9}P_{0.76}Ge_{0.24}$ | $Mn_{1.06}Fe_{0.94}(P_{0.80}Ge_{0.20})$ | 6.7 |
| LL084- 2# | $Mn_{1.1}Fe_{0.9}P_{0.76}Ge_{0.24}$ | $Mn_{1.07}Fe_{0.93}(P_{0.82}Ge_{0.18})$ | 7.7 |
| YM04 | $Mn_{1.1}Fe_{0.9}P_{0.76}Ge_{0.24}$ | $Mn_{1.05}Fe_{0.95}(P_{0.80}Ge_{0.20})$ | 6 |
| YM01- 3# | $Mn_{1.1}Fe_{0.9}P_{0.8}Ge_{0.2}$ | $Mn_{1.06}Fe_{0.94}(P_{0.76}Ge_{0.24})$ | 6.2 |
| YM05 | $Mn_{1.1}Fe_{0.9}P_{0.8}Ge_{0.2}$ | $Mn_{1.06}Fe_{0.94}(P_{0.82}Ge_{0.18})$ | 6.8 |
| YM06 | $Mn_{1.1}Fe_{0.9}P_{0.8}Ge_{0.2}$ | $Mn_{1.05}Fe_{0.95}(P_{0.78}Ge_{0.22})$ | 6.8 |
| YMA11 | $Mn_{1.1}Fe_{0.9}P_{0.8}Ge_{0.2}$ | $Mn_{1.06}Fe_{0.94}(P_{0.76}Ge_{0.24})$ | 6.2 |
| YM02 | $Mn_{1.1}Fe_{0.9}P_{0.78}Ge_{0.22}$ | $Mn_{1.06}Fe_{0.94}(P_{0.77}Ge_{0.23})$ | 7.2 |

Three typical samples with nominal compositions $Mn_{1.2}Fe_{0.8}(P_{0.76}Ge_{0.24})$ (sample #1), $Mn_{1.1}Fe_{0.9}(P_{0.76}Ge_{0.24})$ (samples #2), and $Mn_{1.1}Fe_{0.9}(P_{0.8}Ge_{0.2})$ (samples #3) have been selected to carry out detailed temperature and magnetic field neutron diffraction measurements. The refined compositions for these three samples are #1-$Mn(Fe_{0.83}Mn_{0.17})(P_{0.74}Ge_{0.26})$, #2-$(Mn_{0.973}Fe_{0.027})(Fe_{0.923}Mn_{0.077})(P_{0.83}Ge_{0.17})$, and #3-$(Mn_{0.994}Fe_{0.006})(Fe_{0.938}Mn_{0.062})(P_{0.76}Ge_{0.24})$.

In order to determine the nature of the transition on the composition and detailed crystal and magnetic structures of the three compositions of $Mn_xFe_{2-x}(P_{1-y}Ge_y)$, in addition to the data at 295 K where the materials are in the paramagnetic state, complete NPD data sets have been collected at lower temperature, well within the ferromagnetic state where the spontaneous moment is saturated, and in a high applied magnetic field where the field-induced moment is saturated. These we designate (T-PFT) and (F-PFT) to indicate the condition when the phase has fully transformed with temperature, or field, respectively. The crystallographic refinement results at room temperature are provided in Table 2.

**Table 2.** Structural parameters of $Mn_xFe_{2-x}(P_{1-y}Ge_y)$ at 295 K. Space group: $P\bar{6}2m$. Atomic positions: Fe/Mn(1): $3f(x, 0, 0)$; Mn(2): $3g(x, 0, 1/2)$; P/Ge(1): $1b(0, 0, 1/2)$; P/Ge(2): $2c(1/3, 2/3, 0)$. The presence of few percent MnO impurity phase was detected and taken into account in the final refinements for all the samples.

| Sample number | | 1# | 2# | 3# |
|---|---|---|---|---|
| Nominal composition | | $Mn_{1.2}Fe_{0.8}(P_{0.76}Ge_{0.24})$ | $Mn_{1.1}Fe_{0.9}(P_{0.76}Ge_{0.24})$ | $Mn_{1.1}Fe_{0.9}(P_{0.8}Ge_{0.2})$ |
| Refined composition | | $Mn_{1.17}Fe_{0.83}(P_{0.74}Ge_{0.26})$ | $Mn_{1.05}Fe_{0.95}(P_{0.83}Ge_{0.17})$ | $Mn_{1.06}Fe_{0.94}(P_{0.76}Ge_{0.24})$ |
| MnO impurity [%] | | 3.4(2) | 4.5(1) | 3.4(2) |
| PMP fraction [%] | | 96.6(2) | 95.5(1) | 96.6(2) |
| $a$ (Å) | | 6.10455(9) | 6.06932(5) | 6.0655(1) |
| $c$ (Å) | | 3.44797(6) | 3.45731(4) | 3.46270(6) |
| $V$ (Å$^3$) | | 111.276(4) | 110.294(2) | 110.326(4) |
| Mn(2) 3g | $x$ | 0.5925(3) | 0.5918(3) | 0.5918(3) |
| | $B$(Å$^2$) | 0.88(2) | 0.61(2) | 0.61(2) |
| | $n$(Mn/Fe) | 1/0 | 0.973/0.027(3) | 0.973/0.027(3) |
| Fe/Mn(1) 3f | $x$ | 0.2528(1) | 0.2526(1) | 0.2526(1) |
| | $B$(Å$^2$) | 0.88(2) | 0.82(4) | 0.82(4) |
| | $n$(Mn /Fe) | 0.170(3) / 0.830 | 0.077(5) / 0.923 | 0.077(5) / 0.923 |

| P/Ge(1) | $B(Å^2)$ | 0.78(3) | 0.51(3) | 0.51(3) |
| 1b | n(P/Ge) | 0.895/0.105(17) | 0.976/0.024(19) | 0.976/0.024(19) |
| P/Ge(2) | $B(Å^2)$ | 0.78(3) | 0.68(3) | 0.68(3) |
| 2c | n(P/Ge) | 0.659/0.341(17) | 0.751/0.249(19) | 0.751/0.249(19) |
| $R_p$ (%) | | 4.42 | 4.20 | 4.20 |
| $wR_p$ (%) | | 5.38 | 5.39 | 5.39 |
| $\chi^2$ | | 1.75 | 1.44 | 1.44 |

By comparing samples #1 with #2 we note that increasing the Mn content results in an increase of the lattice parameter *a*, a decrease of *c*, and an increase of the unit cell volume *V*, while, a comparison of samples #2 and #3 shows that with increasing Ge content the *a* axis decreases and *c* axis and *V* increase. The *V* increase can be attributed to the substitution of larger Mn and Ge for the smaller Fe and P, respectively. It is interesting that the *c/a* ratio increases with increasing Ge content, and decreases with increasing Mn content, indicating that the larger Mn and Ge atoms have an opposite effect on the variations of the *a*- and *c*-axes.

### 3.2. Temperature and field induced transition

Thermomagnetic curves *M-T* have been measured for the three samples and are shown in Fig. 2a, and are compared in Fig. 2b with the FM phase fractions as a function of temperature (*frac*(FM)-*T*) derived from monitoring the integrated intensities of the (001) neutron Bragg reflection upon warming and cooling. Note that the *c*-axis lattice parameters in the paramagnetic and ferromagnetic phases are quite different, so that $(001)_{PM}$ and $(001)_{FM}$ are easily distinguishable with NPD [9,10]. The ferromagnetic transition temperature ($T_c$), thermal hysteresis ($\Delta T_{hys}$), and the temperature range of coexistence of the PM and FM phases ($\Delta T_{coex}$) [9,10] are given in the figure and are quite different for the three samples. For each sample, the corresponding values of $T_C$ derived from the *M-T* and Frac(FM)-*T* curves are in good agreement on both warming and cooling. The values of $\Delta T_{hys}$ and $\Delta T_{coex}$ observed in the neutron experiments are larger than those derived in the magnetization measurements. The neutron data (Fig. 2b) show that over this temperature range only ~85% of the PM phase is transformed into the FM phase in all three samples. Data are summarized in Table 3.

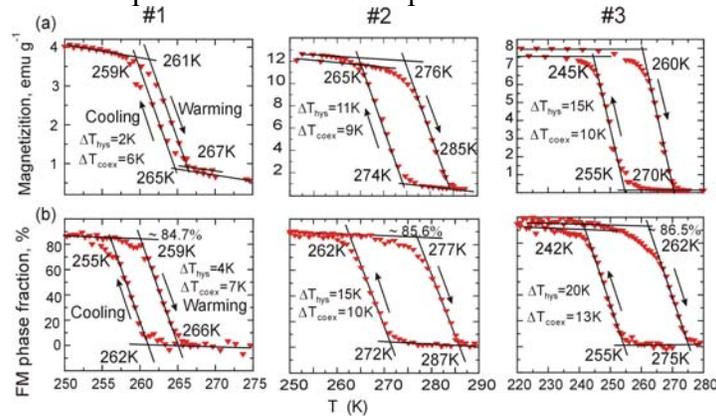

**Figure 2.** (a) (color online) Temperature dependence of magnetization *M-T* of $Mn_xFe_{2-x}(P_{1-y}Ge_y)$ for three polycrystalline compositions (see Table 2) obtained on warming and cooling under a magnetic field of 0.05 T. Note that 1 emu/g = 1 $Am^2$/kg; (b) FM phase fraction as a function of temperature upon warming and cooling under zero applied field.

In order to explore the effects of an applied magnetic field on the PM-FM transition, neutron diffraction data were collected under fields of 0-7 T. As an example, data obtained from sample #1 are shown in Fig. 3, in which the integrated intensity of the (001) reflection of the PM phase is plotted as a function of the magnetic field at different temperatures near $T_C$. Note that only about 85% of the PM phase is transformed into the FM phase at high fields, independent of temperature, in agreement with the results of the temperature-induced PM-FM transition shown in Fig. 2b. The data also indicate that the PM-FM transition takes place above a critical field $B_{on}$ which depends on the temperature. The fields $B_{on}$ and $B_{end}$, defined as the onset (*on*) and completion (*end*) of the transition, were determined at different temperatures by neutron diffraction for a variety of $Mn_xFe_{(2-x)}P_{(1-y)}Ge_y$ samples listed in Table 1. These data were used to construct the magnetic phase diagram shown in the inset of Fig. 3. Between $B_{on}$ and $B_{end}$, the PM and FM phases coexist and the range of coexistence $\Delta B = B_{end} - B_{on}$ is about 1T. The $B_{on}$ and $B_{end}$ curves are almost parallel and increase approximately linearly with increasing temperature at the rate $dB_{on}/dT$ of about 0.3T/K. This indicates that the temperature dependence of $H_{on}$ is strong, which is a favorable property for obtaining high values of $\Delta S_M$ due to the fact that according to the Clausius-Clapeyron relation [14] the magnetic entropy change depends not only on the magnetization jump at $T_C$ but also on the temperature dependence of $H_{on}$.

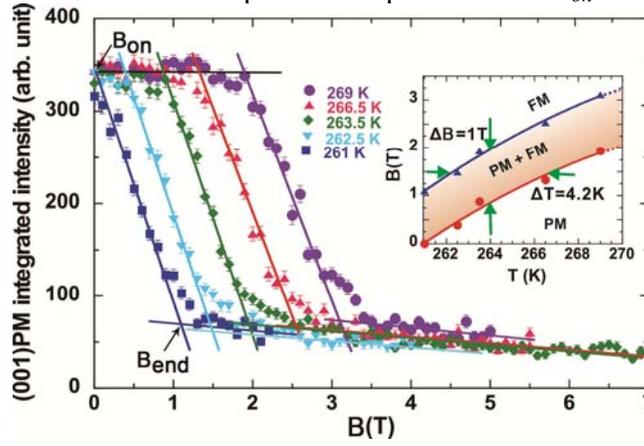

**Figure 3.** (color online) $(001)_{PM}$ integrated intensity of sample #1 as a function of magnetic field at different temperatures. Inset: magnetic phase diagram for sample #1 obtained from neutron data.

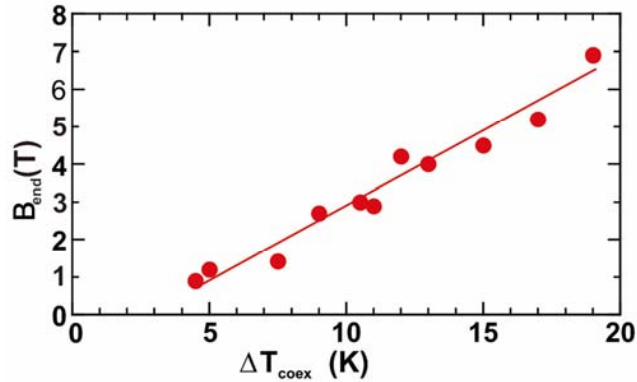

**Figure 4.** (color online) Plot of the $B_{end}$ as a function of $\Delta T_{coex}$.

The plot of $B_{end}$ vs. $\Delta T_{coex}$, shown in Fig. 4, indicates that there is a close correlation between $\Delta T_{coex}$ and $B_{end}$, *i.e.*, the larger is $\Delta T_{coex}$, the larger is $B_{end}$. The data were collected from the 11 $Mn_{1.1}Fe_{0.9}P_{0.8}Ge_{0.2}$, $Mn_{1.1}Fe_{0.9}P_{0.76}Ge_{0.22}$, $Mn_{1.1}Fe_{0.9}P_{0.76}Ge_{0.24}$ or $Mn_{1.2}Fe_{0.8}P_{0.76}Ge_{0.24}$ sintered samples as shown in Table 1. For each sample, we measured the $\Delta T$ on plots of the FM phase fraction as indicated in Fig. 2b, and determined the value of the magnetic field where the transition ended. For example, the $\Delta T_{coex}$ is 7 K (Fig2b) and the magnetic field at which the transition completes is 1 T (Fig. 3a) for sample #1. Repeating this process for all 11 samples, the variation of $\Delta T_{coex}$ as function of the applied magnetic field was obtained. It has been shown that the coexistence of the PM and FM phases in the temperature interval $\Delta T_{coex}$ is due to the inhomogeneity of the distribution of Ge atoms in the structure [9,10]. It follows, therefore, that the system will have small values of $B_{end}$ when the chemical inhomogeneity is reduced or eliminated by means of appropriate preparation and treatment of the samples [9,10].

**3.3. Entropy change determined from calorimetric and magnetic measurements**

Typically the MCE can be derived from magnetic measurements with the use of the Maxwell relation, but the reliability of this procedure in the case of first order transitions is controversial [15,16]. Therefore, for our system, calorimetric measurements made with a differential scanning calorimetry (DSC) have also been made to directly obtain the latent heat and the entropy change [17,11]. The results for the two methods can then be compared.

The PM-FM transition can be temperature-induced, on cooling under zero-field, and the DSC technique can be used to obtain the zero-field heat capacity data. The total entropy $S_{H=0}(T)$ can then be evaluated by numerical integration of the expression

$$S(T,H) = \int_0^T \frac{C_p(T,H)}{T} dT + S_0 \qquad (1)$$

where $S_0$ is the entropy at $T = 0$ K (generally assumed to be zero), and $C_p$ is the zero-field heat capacity. Here, for convenience, S(T,0) for $B$=0 is renamed as $S_{DSC}$.

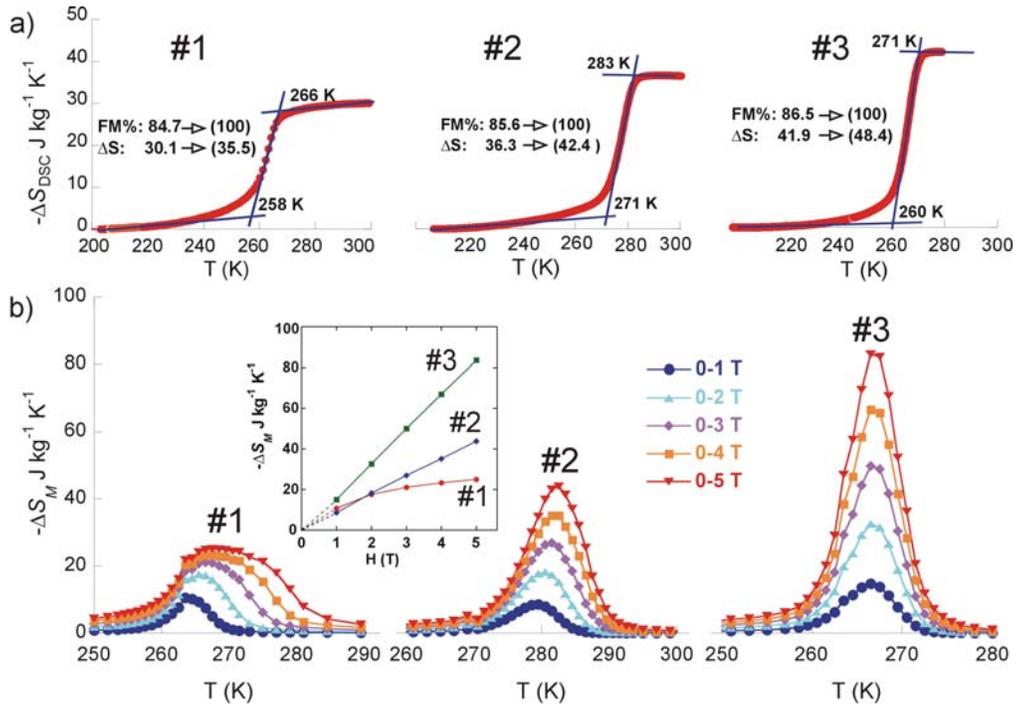

**Figure 5.** (color online) a) Zero field entropy changes as a function of temperature obtained from DSC data taken with 5 K/min; b) Temperature dependence of the magnetic entropy change of the three $Mn_xFe_{2-x}(P_{1-y}Ge_y)$ samples as a function of applied magnetic field determined using the Maxwell relation. The inset shows the peak $|\Delta S_M|$ as function of field at $T_C$ for the three samples.

Fig. 5a shows $\Delta S_{DSC}$ obtained from the DSC data for the three $Mn_xFe_{2-x}(P_{1-y}Ge_y)$ samples as a function of temperature at rates of 5 K/min on warming. The transition temperatures agree very well with the results obtained from the *M-T* data and from the neutron measurements (Fig. 2). The directly measured entropy changes $\Delta S_{DSC}$ obtained for samples #1, #2, and #3 are 30.1, 36.3, and 41.9 J kg$^{-1}$ K$^{-1}$, respectively, and they include both the magnetic $|\Delta S_M|$ and the structural ($|\Delta S_{lat}|$) contributions of the first order transition. These values can be corrected for the effect of the incompleteness of the PM-FM transition and for the small MnO impurity content, yielding normalized values of $\Delta S_{DSC}$ of 35.5, 42.4, and 48.4 J kg$^{-1}$ K$^{-1}$ for $Mn(Fe_{0.83}Mn_{0.17})(P_{0.74}Ge_{0.26})$-#1, $(Mn_{0.973}Fe_{0.027})(Fe_{0.923}Mn_{0.077})(P_{0.83}Ge_{0.17})$-#2, and $(Mn_{0.994}Fe_{0.006})(Fe_{0.938}Mn_{0.062})(P_{0.76}Ge_{0.24})$-#3, respectively, assuming metallurgically ideal samples. These are the values given in Table 3 and those used in the following discussions.

**Table 3.** Structural parameters and magnetic properties of $Mn_xFe_{2-x}(P_{1-y}Ge_y)$. The ferromagnetic structure has magnetic symmetry *P11m'* with moments for Mn and Fe that lie within the *a-b* plane. For convenience the moment was set parallel to the *a* direction in the refinements; the moment direction within the *a-b* plane cannot be determined by NPD.

| Sample | #1, *x*=1.17, *y*=0.26 | #2, *x*=1.05, *y*=0.17 | #3, *x*=1.06, *y*=0.24 |
|---|---|---|---|
| *n*%Mn at 3*f* site at 295 K | 17.0(3) | 7.7(5) | 6.2(4) |
| *c/a* of PM at 295 K | 0.56482(1) | 0.56964(1) | 0.57088(1) |
| *c/a* of FM after T-PFT, 0 T | 0.54568(4), at 250K<br><br>0.5502, at 262 K | 0.54544(2), at 259K | 0.54384(2), at 230K |
| *c/a* of FM after F-PFT | 0.54714(4), 262K/3T | 0.54673(2), 271K/3T | 0.54600(2), 253K/5T |
| $\Delta c/a(T)$ in T-PFT [%], 0 T | 1.91(1), 295K~250K<br><br>1.46, 295 K~262K | 2.41(1), 295k ~259K | 2.70(1), 295k ~230K |
| $\Delta c/a(H)$ in F-PFT [%] | 1.77(1),<br><br>295K -262K/0-3T | 2.29(1),<br><br>295K -271K/0-3T | 2.49(1),<br><br>295K -253K/0-5T |
| $M_{Fe}(T)$ [$\mu_B$] (in T-PFT), 0 T | 0.35(7), at 250K | 0.73(6), at 259 K | 1.14(9), at 230 K |
| $M_{Mn}(T)$ [$\mu_B$] (in T-PFT), 0 T | 3.15(7), at 250K | 3.56(5), at 259 K | 3.57(9), at 230 K |
| $M_{f.u}(T)$ [$\mu_B$] (in T-PFT), 0 T | 3.50(10), at 250K | 4.30(8), at 259 K | 4.7(1), at 230 K |
| $M_{Fe}(B)$ [$\mu_B$] (in F-PFT) | 0.89(9), 262K/0-3T | 0.75(8), 271K/0-3T | 0.80(7), 253K/0-5T |
| $M_{Mn}(B)$ [$\mu_B$] (in F-PFT) | 3.70(9), 262K/0-3T | 3.91(8), 271K/0-3T | 4.49(8), 253K/0-5T |

| $M_{f.u}(B)$ [$\mu_B$] (in F-PFT) | 4.59(10), 262K/0-3T  4.30(1), 262K/0-1T | 4.64(10), 271K/0-3T  4.72(10), 271K/0-6T | 5.5(1), 253K/0-3T  5.6(1), 253K/0-5T |
|---|---|---|---|
| *$\Delta S_{DSC}$ J Kg$^{-1}$K$^{-1}$, 0 T | 35.5 | 42.4 | 48.4 |
| **$\Delta S_M, calc$ J kg$^{-1}$K$^{-1}$ | 46.6 (3T), 43.6 (1 T) | 45.8 (3T), 46.5 (6T) | 56.6 (3T), 57.7 (5T) |
| $T_C$ [K] | 267 | 285 | 270 |
| $\Delta T_{hys}$ [K] | 2 | 11 | 15 |
| $\Delta T_{coex}$ [K]/$\Delta H$(T) | 6/1.2 | 9/2.7 | 10/3.0 |
| $n_{rec}$%Unchanged PM[%] | 15.3 at 250 K | 14.4 at 259 K | 13.5 at 230 K |

*Normalized values $\Delta S_{DSC}$ obtained from Fig. 5a.
** $\Delta S_M, calc = M_{f.u}(B) \times [\Delta S_{DSC} / M_{f.u}(T)]$.

In Fig. 5b we show the curves |$\Delta S_M$| for the three samples as a function of temperature and for different magnetic fields, in order to compare the values of the entropy changes obtained in this work with those reported for different magnetocaloric materials and evaluated by the same method [1-7,18]. Determination of the isothermal $\Delta S_M$ from *M-H* data is commonly done using the Maxwell equation. The calculated maximum are |$\Delta S_M$| = 25.1, 43.8, and 83.5 J kg$^{-1}$K$^{-1}$ for samples #1, #2, and #3, respectively. When normalized assuming ideal properties, these values increase to 29.6, 51.2, and 96.5 J kg$^{-1}$ K$^{-1}$. The inset of Fig. 5b shows the peak |$\Delta S_M$| as a function of field at $T_C$. For samples #2 and #3, |$\Delta S_M$| increases linearly with increasing field at a rate of 9.7, and 20.2 J kg$^{-1}$K$^{-1}$H$^{-1}$, respectively, while for sample #1 the rate is 8.4 J kg$^{-1}$K$^{-1}$H$^{-1}$ below 2 T, and only 1.4 J kg$^{-1}$K$^{-1}$H$^{-1}$ above 2 T. Since the critical field for the transition is 2 T, as mentioned above, this behavior indicates that |$\Delta S_M$| of sample #1 is nearly saturated under ~2 T at $T_C$. The larger fields broaden the temperature plateau instead of enhancing the entropy change, which corresponds to a shift of the critical temperature of the PM-FM transition, as shown in the inset of Fig. 3. Thus the peak of |$\Delta S_M$| of sample #1 exists in a temperature range (from 255 up to 275 K) wider than those of samples #2 and #3. In fact, the peaks of |$\Delta S_M$| of samples #2 and #3 do not shift significantly to higher temperatures with increasing field.

### 3.4. Relationships between n%Mn at the 3f site and physical properties

The PM and FM phases in all the compounds have the same crystal symmetry ($P\bar{6}2m$) and structure, and therefore we expect that the observed variations of the physical properties must be strongly correlated with the chemical composition and the distribution of the atoms in the structure. Table 3 only gives the structural parameters and magnetic properties of the three Mn$_x$Fe$_{2-x}$(P$_{1-y}$Ge$_y$) samples, 1#, 2# and 3#. Fig. 6 shows such data for all 11 compounds. Indeed the data reported in Table 3 and Fig. 6 show that the substitution of Mn for Fe atoms in the 3*f* site (*n*%Mn-3*f*) plays a very important role in this system.

From Table 1 we found that the *n*%Mn-3*f* value depends on the Mn/Fe but not on the Ge/P ratio in the alloys. As the concentration of Mn atoms located in the 3f sites changes, the parameters corresponding to the most important physical properties are significantly affected, as shown in Table 3 and Fig. 6. We note that in discussing the values of $\Delta S_{DSC}$, the direct calorimetric measurements made with a differential scanning calorimetry (DSC) will be used rather than the values obtained with the Maxwell relation using the magnetization curves.

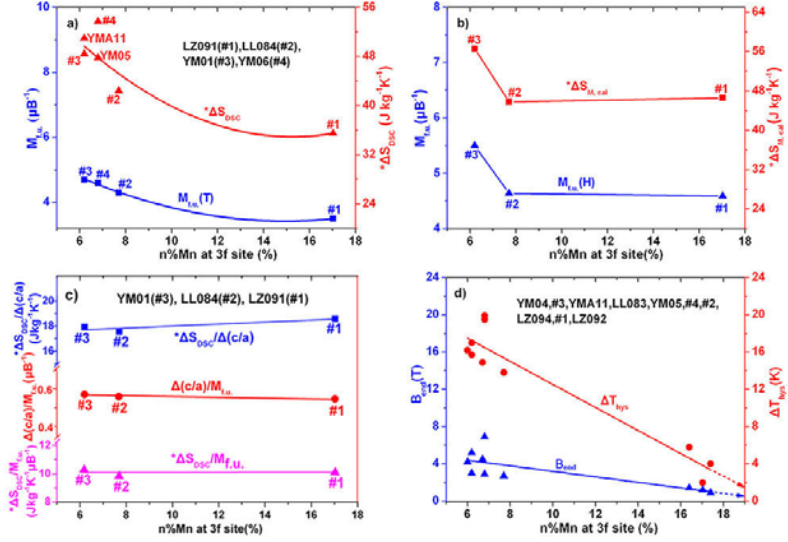

Figure 6. (color online) Variation of some physical properties of the compounds of the system $Mn_{2-x}Fe_xP_{1-y}Ge_y$ as a function of the atomic per cent of Mn in the 3f position of space group $P\bar{6}2m$: a) total magnetic moments $M_{f.u.}(T)$, $M_{Mn}(T)+M_{Fe}(T)$ of the FM phase, obtained from the temperature dependent NPD, and $*\Delta S_{DSC}$ normalized and calculated (The 'normalized' entropy change is from the $\Delta S_{DSC}$ obtained by extrapolating, from the measured value obtained from the incomplete transition, the value obtainable assuming a 100% conversion of PM to FM); b) total magnetic moments $M_{f.u.}(H)$, $M_{Mn}(B)+M_{Fe}(B)$, of the FM phase, obtained with an applied magnetic field of 1 and 3T from NPD measurements. The 'calculated' $*\Delta S$ is obtained by $*\Delta S_{CALC} = M_{f.u.}(B)*\Delta S_{DSC}/M_{f.u.}(T)$; c) this figure shows that the changes of $*\Delta S_{DSC}$ and $\Delta(c/a)$ as a function of n% are related by the equation $*\Delta S_{DSC} = k \times \Delta(c/a)$ where k is a constant. Similar behavior is found for $*\Delta S_{DSC}$ and the total magnetic moment $M_{Fe}(T)$ (see also Fig. 7). It follows that $\Delta(c/a)/M_{f.u.}(T) = k'/k$ where $*\Delta S_{DSC}/M_{f.u.}(T) = k'$. The ratio k'/k is slightly sloping down as n% increases; d) the values of $B_{end}$ (intensity of the magnetic field at the end of the transition) and $\Delta T_{hys}$ as function of n%. Note that $\Delta T_{hys}$ decreases sharply as $n\%$ increases.

More specifically, when $n\%$Mn-$3f$ increases from 6.2% in sample #3 to 17% in sample #1, the magnetic moments decrease from 4.7 to 3.5 $\mu_B$ for $M_{f.u.}(T)$ and from 5.5 to 4.59 $\mu_B$ $M_{f.u.}(B)$, with a consequent decrease in the normalized $*\Delta S_{DSC}$ from 48.4 to 35.5 J kg$^{-1}$ K$^{-1}$ (Fig. 6a). Most importantly, there is a huge decrease of the thermal hysteresis from 15 to 2 K, and a huge decrease of the magnetic field saturation requirement from 7 to 0.9 T (Fig.6d). Fig. 6c shows that the ratios of $*\Delta S_{DSC}/M_{f.u}(T)$, $\Delta c/a/M_{f.u}(T)$ and $*\Delta S_{DSC}/\Delta c/a$ vs. $n\%$Mn-$3f$ are almost constant, and that the ratio $*\Delta S_{DSC}/M_{f.u.}(T)$ has a constant value of about 10 J kg$^{-1}$ K$^{-1}$$\mu_B^{-1}$ for the three samples. This behavior suggests that the total entropy change and the magnetostriction effect are linearly correlated with the ordered magnetic moment. The function $*\Delta S_{DSC}$ vs. $M_{f.u}(T)$ is plotted in the lower part of Fig. 7. The figure also shows that $\Delta S_M$, evaluated using the equation $\Delta S_M^{max}=R\ln(2J+1)$ [19] vs. the moment is also almost linear in the range between 2 and 6 $\mu_B$. On the basis of these results, a compound like sample #1 with ~19%Mn-$3f$ is expected to have a large MCE ($\Delta S_{M, calc}$ in Table 3 and Fig. 5a), between 35 and 46 J kg$^{-1}$ K$^{-1}$, with small hysteresis ($\Delta T_{hys}$ close to 0 K) and a required magnetic field of less than 1.2 T (Fig. 5c).

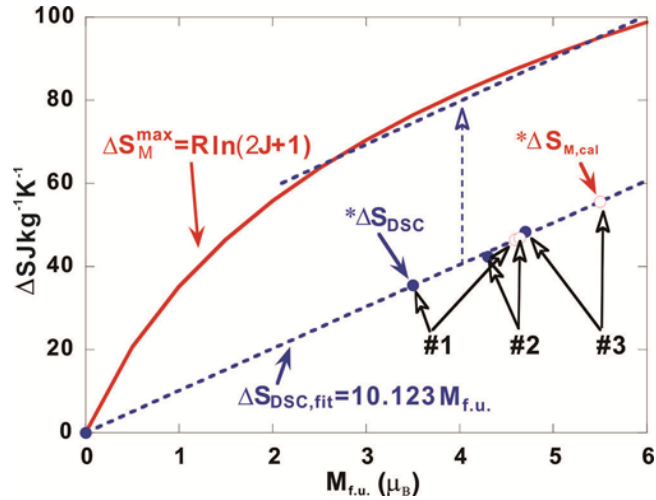

Figure 7. (color online) Plot of $\Delta S_M^{max}$, $^*\Delta S_{DSC}$ and $^*\Delta S_M$ calc vs. the total magnetic moment $M_{f.u.}$ (T). $^*\Delta S_M$ calc was calculated with the equation $^*\Delta S_M\, calc = M_{f.u}(B) \times [^*\Delta S_{DSC} / M_{f.u}(T)]$. Note that it is almost linear in the range between 2 $\mu_B$ and 6$\mu_B$.

In order to have a large MCE with a low applied magnetic field, the material must have a sharp first-order transition with a large entropy change from one state to the other, and large ordered magnetic moments. It is also important to have a small thermal hysteresis to reduce losses, and a small magnetostriction for mechanical stability. The hysteresis will in fact reduce the magnetocaloric working efficiency. However, for sample #1, the reduction of $M_{f.u.}(T)$ can be circumvented by using an applied magnetic field of 1 T, with a consequent change of the moment from 3.5 to 4.3 $\mu_B$ and a corresponding increase of MCE from 35.5 to 43.6 J kg$^{-1}$ K$^{-1}$. The small hysteresis for sample #1 can be attributed to the large substitution of Mn for Fe at the 3f site and that of Ge for P. This is because 1) the larger the substitution, the easier chemical homogeneity can be obtained. and 2) the Ge substitution for P increases $T_C$ while Mn substitution for Fe decreases $T_C$, and as a consequence there is little change of $T_C$ [6, 7] in every crystallite particle in the sample. Thus, the magnetic field required for the PM-FM transition is reduced.

The study of a number of samples with different compositions (Table 3 and Fig. 6) has shown that when $n(Mn)\%3f$ increases, a decrease is observed of the following parameters associated with the transition: (i) total magnetic moment ($M_{f.u.} = M_{Mn} + M_{Fe}$; (ii) entropy change $^*\Delta S$; (iii) hysteresis $\Delta T_{hys.}$; (iv) intensity of the applied magnetic field at the end of transition $B_{end}$; (v) magnetostriction $\Delta(c/a)$; (vi) critical temperature $T_C$. Some of these variations are desirable (e.g. hysteresis, $B_{end}$, $\Delta(c/a)$, $T_C$) others are not (e.g. magnetic moment, entropy change ), and, therefore, compromises will have to be made to obtain a compound suitable for practical applications. Fig. 6c shows that the variation of $^*\Delta S_{DSC}$ and of $\Delta(c/a)$ as a function of $n(Mn)\%3f$ is related by the equation $^*\Delta S_{DSC} = k \times \Delta(c/a)$, where k is constant. Similar behavior is found for $^*\Delta S_{DSC}$ and the total magnetic moment $M_{f.u}(T)$ (see Fig.7). It therefore follows that $\Delta(c/a)/M_{f.u.}(T) = k'/k$, where k'= $^*\Delta S_{DSC} / M_{f.u}(T)$ (in fact k'/k is very slightly sloping down as n% increases). Two of the most important properties that a magnetocaloric material must have to be a candidate for technological applications are a sharp hysteresis of the PM to FM transitions and a small magnetic field necessary to induce the transition around room temperature. As shown in Fig. 6d, both are strongly dependent on $n(Mn)\%3f$, making the Mn content of a compound of the $Mn_xFe_{2-x}P_{1-y}Ge_y$ system one of the primary factors to consider when planning to prepare materials suitable for practical applications.

**3.5. Effects of crystallite size and inhomogeneities of composition on the transition**

As shown in Fig. 2b and 3, the transition PM-FM is not complete, and it has been observed that the diffraction lines of the residual PM phase are considerably broadened, indicating that the size of the PM crystallites that don't transform is small.

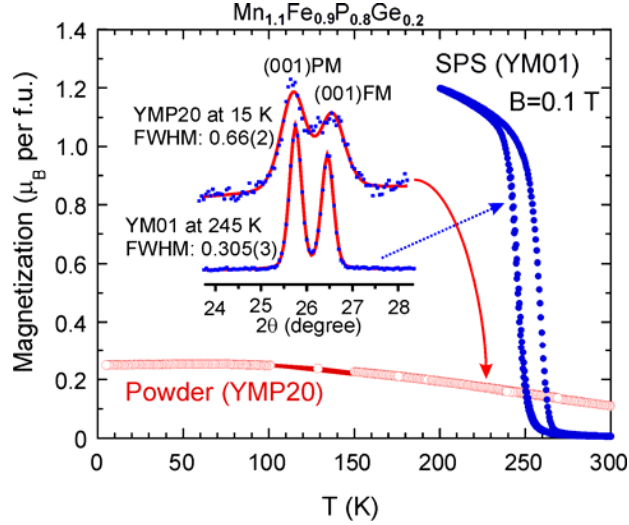

Figure 8. (color online) Magnetization vs temperature plots of the Ball Milled (BM) and Spark Plasma Sintering (SPS) samples (YMP20 and YM01, respectively) obtained from the compound having nominal composition $Mn_{1.1}Fe_{0.9}P_{0.8}Ge_{0.2}$. The neutron diffraction profiles of the (001)PM and (001)FM reflections produced by the two samples are shown in the inset. The results demonstrate that in the case of YMP20 (very broadened peaks) the transition from PM to FM is hindered because the crystallites are too small to allow the ordering of the magnetic moments, while in the case of YM01 (sharp peaks) the transition occurs as expected.

Fig. 8 shows the magnetization curves of the powder (YMP20) and SPS synthesis (YM01) samples of $Mn_{1.1}Fe_{0.9}P_{0.8}Ge_{0.2}$ as a function of temperature. The YMP20 sample was prepared by ball milling while the YM01 sample was prepared by spark plasma sintering. A ferromagnetic transition occurs near 250 K for the YM01 sample but no sharp transformation was observed in the measurement for the YMP20 powder. The insert shows that the primary difference between the two sample is that the width of the 001 PM and FM reflections is much larger in the case of YMP20 (at 15 K) than that of YM01 (at 245 K). Correspondingly, the magnetization curve of YM01 is sharp, and that of YMP20 remains practically unchanged between 4 and 300 K. This indicates that the small sizes of the crystallites (estimated from the width of the (001) Bragg peaks to be about 50 nm for YMP20) inhibit the transition. These results demonstrate that the crystallized particle size plays an essential role in the magnetocaloric properties.

Fig. 9a displays the zero-field temperature dependence of the intensities and widths of the 001 PM and FM peaks for the YM01 $Mn_{1.1}Fe_{0.9}P_{0.8}Ge_{0.2}$ sintered sample, collected on warming at 10 K/hour. The inset shows the diffraction scan (2θ) of the 001 PM and FM peaks at the indicated temperatures. At 200 K, a fraction of the sample is still in the PM phase. This fraction clearly consists of very small crystallites, as indicated by the large width of the 001 PM peak. The contribution of the small crystallites to the width of the 001 PM peak begins to be significant at 265 K and keeps increasing down to 200 K. Note that the width of the FM peak is constant over the entire range of temperatures. The phase transition occurs in the interval 258-275 K ($\Delta T = 17$ K), in which the intensity changes with a steep slope, and, at a much slower rate, in the interval 248-258 K. The slope in the range 258-275 K indicates inhomogeneities of composition, and the small departures from linearity are due to the fact that the compositional variations in the sample are not quite uniform.

Fig. 9c,d displays the magnetic field dependence of the intensities and widths of the 001 PM and FM peaks for the YM01 $Mn_{1.1}Fe_{0.9}P_{0.8}Ge_{0.2}$ sintered sample at 255 K, the PM-FM Curie temperature $T_C$. Between 0 and 2.6 T the intensities of the two peaks vary slowly, between 2.6 T and 5.5 T the variation is more pronounced, and finally between 5.5 T and 6.9 T the remaining PM phase essentially does not convert into FM. Interestingly, the FWHM of (001) PM increases in this last interval significantly, confirming once again that the smallest crystallites in the sample are responsible for the incompleteness of the transformation.

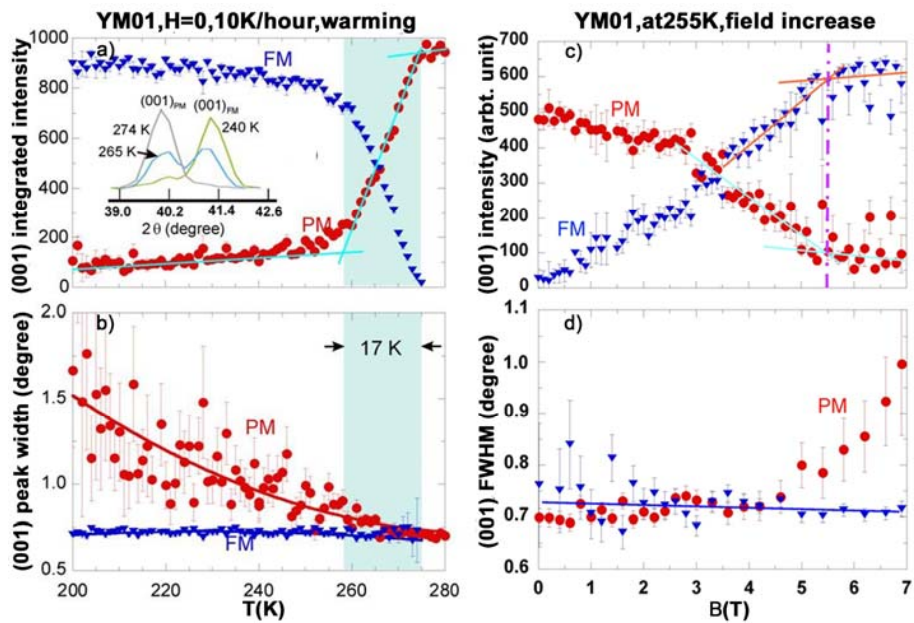

Figure 9. (color online) Neutron diffraction intensities of the (001) PM and FM reflections as function of: a) temperature and c) applied magnetic field for the YM01 sample. The corresponding linewidths are shown in b) and d), respectively. The figure illustrates that in both the temperature and magnetic field induced transitions: (i) the transition is not complete, (ii) the ranges of coexistence of the two phases are large and, (iii) the broadening of both the FM and PM peaks at the beginning and at the end of the transition is quite evident. As explained in the text, these features are due to the distribution of crystallite sizes in the sample. The inset shows the profiles of the (001) Bragg peak in the PM and FM phases at three temperatures.

A comparison of Fig. 9c with Fig. 9a indicates that the effect of an applied magnetic field is similar to that produced by lowering the temperature, as it should be, since both induce the ferromagnetic ordering of the spins of the magnetic atoms. The two methods, however, are not identical; the changes of intensity and FWHM of the (001) PM reflections are not as well defined in the case of the applied field. One reason for this behavior is that the moments prefer to lie in-plane, while the applied field in a powder makes all angles with respect to the easy plane. Conversely, when lowering the temperature the effect of the internally generated in-plane field assists the external field in facilitating the PM to FM transition.

It is clear that the size of the crystallites plays an important role in the fraction of the sample that transforms, as the data in Fig. 9 demonstrate that the largest crystallites transform first, leaving the smaller crystallites in the PM phase; hence the broadened Bragg peak. In addition, we know from previous work that compositional variations affect both the transition temperature and phase fraction that converts [8-11]. We have therefore undertaken additional characterization and annealing studies in order to unravel these two contributions to the performance of the material and to establish a procedure to improve the performance.

The distribution of compositions in these $Mn_xFe_{2-x}P_{1-y}Ge_y$ alloys was investigated by SEM+EDAX and scanning transmission synchrotron radiation X-ray microscope experiments. The latter X-ray measurements were carried out at the Shanghai Synchrotron Radiation Facility [21], which provided chemical mapping with a spatial resolution of 30 nm. Mn, Fe and Ge were chosen to be scanned separately at the absorption edge and then distributions of these elements were obtained. The element mappings do not show significant inhomogeneity.

SEM+EDAX analysis shows that in our $Mn_xFe_{2-x}P_{1-y}Ge_y$ alloys, Ge readily accumulates at grain boundaries. Fig. 10a shows the backscattered electron SEM image of the DMD01 $Mn_{1.2}Fe_{0.8}P_{0.76}Ge_{0.24}$ bulk sample. The white lines located at the grain boundaries were analyzed by EDAX which indicated that those are areas with enhanced Ge content.

In order to ascertain the effects of both particle size and inhomogeneities on the magnetocaloric properties, subsequent homogenizing treatments were applied to the samples. After a series of systematic heat treatments, it was determined that annealing at 950ºC for 15 h, followed by annealing at 850ºC for 48 h, was the optimum processing parameters. After such heat treatments, both the compositional homogeneity improved and the grains increased in size, substantially improving the magnetocaloric properties as we now discuss.

The DMD01 $Mn_{1.2}Fe_{0.8}P_{0.76}Ge_{0.24}$ sample was annealed using the above procedure, and we designate this DMD02. Fig. 10b shows the EBSD image of the DMD02 $Mn_{1.2}Fe_{0.8}P_{0.76}Ge_{0.24}$ sample, comparing it to the sample before annealing (Fig. 9a). Note that the Ge concentration at the grain boundaries has been obviously reduced. Table 4 gives the atomic distributions at four points (E, F, G, H) calculated from the EDAX analysis. Note in particular that the data at F indicate that the grain boundaries in the DM01 sample are rich in Ge and poor in P.

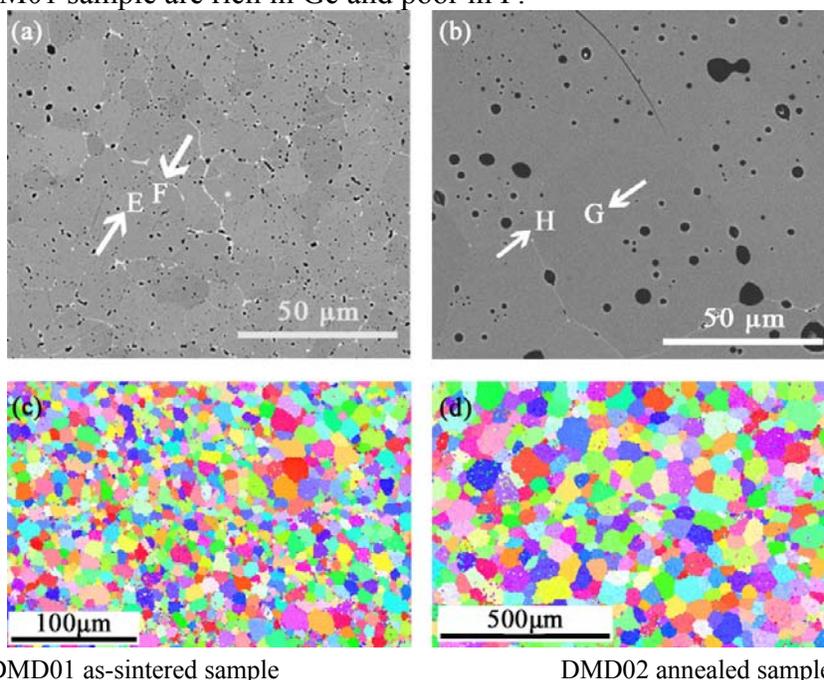

DMD01 as-sintered sample　　　　　　　　　　DMD02 annealed sample

Figure 10. (color online) Backscattered scanning electron microscopy (SEM-BSE) images (a,b) and electron back scattered diffraction (EBSD) images (c,d) of the compound $Mn_{1.2}Fe_{0.8}P_{0.76}Ge_{0.24}$ for a,c) the initial sintered, and b,d) post-annealed samples. The white lines in a) indicate grain boundaries that are rich in Ge. The arrows indicate the points where the SEM-BSE analysis was performed (E and G in-grain points, and F and H grain-boundary points). The black spots in (a,b) are holes. The different shades (colors) appearing in the EBSD images (c,d) are due to the different orientations of the crystallites.

After the above high temperature annealing procedure we find that the grain size has increased dramatically as indicated in the EBSD images in Fig.10(c) and (d) for the DMD01 as-sintered and DMD02 annealed sample. The average grain sizes of the two samples are 10 and 50 μm, respectively, increasing by a factor of five.

Table 4 Atomic distributions of the DMD01 as-sintered and DMD02 annealed $Mn_{1.2}Fe_{0.8}P_{0.76}Ge_{0.24}$ sample

|  | Mn | Fe | P | Ge |
|---|---|---|---|---|
| E, in grain, DMD01 | 1.2096 | 0.7903 | 0.7837 | 0.2291 |
| F, grain boundary, DMD01 | 1.0978 | 0.9463 | 0.5296 | 1.1296 |
| G, in grain, DMD02 | 1.2061 | 0.8015 | 0.7874 | 0.2323 |
| H, grain boundary, DMD02 | 1.2043 | 0.7982 | 0.7786 | 0.2258 |

Fig. 11 presents the $\Delta S_{DSC}$-$T$ curves, and the temperature dependence of the content of the paramagnetic phase percentage calculated from X-ray diffraction for the DMD01 as-sintered and DMD02 annealed $Mn_{1.2}Fe_{0.8}P_{0.76}Ge_{0.24}$ alloys. The transition range $\Delta T$ of the DMD01 and DMD02 $Mn_{1.2}Fe_{0.9}P_{0.76}Ge_{0.24}$ samples is 8.4 K and 6.7 K, respectively, while the residual PM phase at the end temperatures of the transition is 21% and 11.8%. Our previous study demonstrated that the magnetic-entropy change directly corresponds to the phase fraction of the magnetic transition [9], where more phase fraction means larger entropy change. Fig. 11a shows that the $\Delta S_M$ values for DMD01 and DMD02 are 20.5 J/kg·K and 23.1 J/kg·K, respectively, while the thermal hysteresis of DMD02 was reduced from 3.8 K to 3.4 K. Thus the magnetocaloric properties are substantially improved in all aspects by the above annealing procedures.

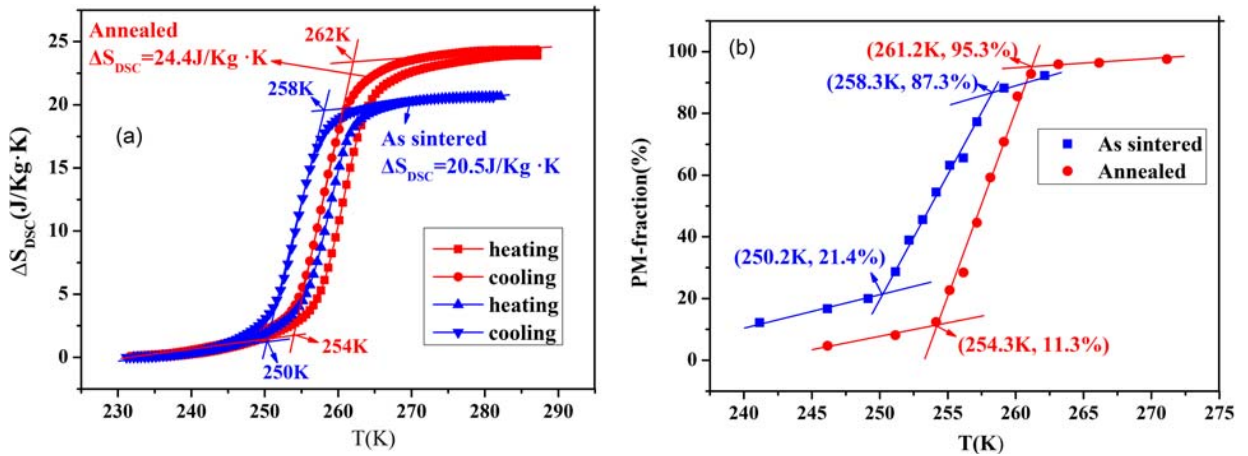

Fig.11. (color online) (a) $\Delta S_{DSC}$ vs temperature plots. The (blue) triangles represent the behavior of the DMD01 sample (as sintered) and the (red) square/circles that of DMD02 (annealed). The increase in the value of $\Delta S$ for the annealed sample is due to the increase of the FM phase fraction in DMD02. (b) phase fraction for the sintered [(blue) squares] and annealed [(red) circles] samples vs. temperature.

## 5. Conclusions

The magnetocaloric properties of the $Mn_xFe_{2-x}P_{1-y}Ge_y$ system near optimal doping have been investigated in detail using a variety of experimental techniques to determine the crystallographic structure, caloric behavior, and morphology of the system. We find that compositional homogeneity and particle size play an important role in optimizing the MCE properties, and an annealing procedure is obtained to optimize these properties.

Based on neutron diffraction, differential scanning calorimetry, and magnetization measurements, we have analyzed the effects of the Mn content and crystallographic location, and of Ge content, on the crystal structure, the magnetic moment, the magnetic entropy change, hysteresis, and saturation field of the first-order PM-FM combined magnetic and structural transition in the system $Mn_{(x)}Fe_{(2-x)}P_{(1-y)}Ge_{(y)}$. The results show that when $n$%Mn-$3f$ increases from 6.2% to 17%, the magnetic moments decrease from 4.7 to 3.41 $\mu_B$ for the temperature induced moment $M_{f.u.}(T)$, and from 5.5 to 4.59 $\mu_B$ for the magnetic field induced moment, $M_{f.u.}(B)$, with a consequent decrease of the entropy change *$\Delta S_{DSC}$ (measured by DSC) from 48.4 to 35.5 J kg$^{-1}$ K$^{-1}$. At the same time, the hysteresis decreases dramatically from 15 K to 2 K, and the value of the applied magnetic field required to complete the PM-FM transition at $T_C$ decreases radically from 7 T to 0.9 T. This behavior indicates that we can control the important properties of these materials by tuning their composition and preparation methods, providing a pathway to optimize those properties relevant for refrigerant applications. We believe that these conclusions may also apply to other compounds with the $Fe_2P$-type structure. We also reported the important correlation between range of temperatures in which the PM and FM phases coexist ($\Delta T_{coex}$) and applied magnetic field which the transition needs to complete ($B_{end}$).

The results make it clear that, in the $Mn_xFe_{2-x}(P_{1-y}Ge_y)$ system, the substitution of Mn for Fe at the $3f$ site causes many changes in the physical properties of the material, and controlling the composition makes it possible to reach a compromise and obtain optimal parameters required to have a good magnetocaloric refrigerant. The compound $Mn(Fe_{0.83}Mn_{0.17})(P_{0.74}Ge_{0.26})$ (sample #1) satisfies all the requirements expected for a good refrigerant, i.e. it has a giant MCE (35.5< MCE<46.5 J kg$^{-1}$ K$^{-1}$), a low magnetic field requirement $B_{end}$ (<1.2 T), small thermal hysteresis $\Delta T_{hys}$ (<2 K), and a reduced (~40%) magnetostriction effect. It is, therefore, a suitable candidate for room temperature magnetic refrigeration applications.

## Acknowledgements

The authors are grateful to Anthony Santoro for valuable discussions and the BL08U Beam line of the Shanghai Synchrotron Radiation Facility (SSRF) for the synchrotron radiation X-ray measurements. This work supported in part by the National Natural Science Foundation of China (51071007, 50731007), the National Basic Research Program of China (2010CB8331001), the Beijing Natural Science Foundation (1112005) and the Knowledge Innovation Project of the Chinese Academy of Sciences. The identification any commercial products does not imply endorsement or recommendation by the National Institute of Standards and Technology.

# References

[1] V. K. Pecharsky, K. A. Gschneider Jr. Phys. Rev. Lett. **78**, 4494 (1997).
[2] O. Tegus, E. Brück, K. H. J. Buschow, R. R. de Boer, Nature **415**, 150 (2002).
[3] F. X. Hu, B. B. Shen, J. R. Sun, Z. H. Cheng, G. H. Rao, X. X. Zhang. Appl. Phys. Lett. **78**, 3675 (2001).
[4] W. Dagula, O. Tegus, X. W. Li, L. Song, E. Brück, D. T. Cam Thanh, F. R. de Boer, K. H. J. Buschow. J. Appl. Phys. **99**, 105 (2006).
[5] D. T. Cam Thanh, E. Brück, O. Tegus, J. C. P. Klaasse, T. J. Gortenmulder, K. H. J. Buschow, J. Appl. Phys. **99**, 08Q107 (2006).
[6] Z. Q. Ou, G. F. Wang, L. Song, O. Tegus, E. Brück, K. H. J. Buschow, Condens. Matter. **18**, 11577 (2006).
[7] N. T. Trung, Z. Q. Ou, T. J. Gortenmulder, O. Tegus, K. H. J. Buschow, E. Brück, Appl. Phys. Lett. **94**, 102513 (2009).
[8] J.V. Leitão. M. van der Haar, A. Lefering, E. Brück J. Magn. Magn. Mater. **344**, 49 (2013).
[9] D. M. Liu, M. Yue, J. X. Zhang, T. M. McQueen, J. W. Lynn, X. L. Wang, Y. Chen, J. Li, R. J. Cava, X. Liu, Z. Altounian, Q. Huang Q, Phys. Rev. B **79**, 014435 (2009).
[10] D. M. Liu, Q. Z. Huang, M. Yue, J. W. Lynn, L. J. Liu, Y. Chen, Z. H. Wu, J. X. Zhang, Phys. Rev. B **80**, 174415 (2009).
[11] Ming Yue, Danmin Liu, Qingzhen Huang, Tong Wang, Fengxia Hu, Jingbo Li, Guanghui Rao, Baogen Shen, Jeffery W. Lynn, and Jiuxing Zhang. J. Appl. Phys. **113**, 043925 (2013).
[12] M. Yue, Z. Q. Li, X. L. Wang, D. M. Liu, J. X. Zhang, and X. B. Liu, J. Appl. Phys. **105**, 07A915 (2009).
[13] A. Larson and R. B. Von Dreele, GSAS: Generalized Structure Analysis System (1994).
[14] A. Giguère, M. Foldeaki, B. Ravi Gopal, R. Chahine, T. K. Bose, A. Frydman, J. A. Barclay, Phys. Rev. Lett. **83**, 2262 (1999).
[15] J. D. Zhou, B. G. Shen, B. Gao, J. Shen, J. R. Sun, Adv. Mater. **21**, 693 (2009).
[16] G. J. Liu, J. R. Sun, J. Shen, B. Gao, H. W. Zhang, F. X. Hu, B. G. Shen, Appl. Phys. Lett. 90, 032507 (2007).
[17] F. Casanova, X. Batlle, A. Labarta, J. Marcos, L. Mañosa, A. Planes, Phys. Rev. B **66**, 212402 (2002).
[18] B. G. Shen, J. R. Sun, F. X. Hu, H. W. Zhang, and Z. H. Cheng, Adv. Mater. **21**, 4545 (2009).
[19] A. M. Tishin, Y. I. Spichkin, Institute of Physics, Bristol and Philadelphia; 2003.1st ed., Vol. 1, Chap. 11, p.351.
[20] D. T. C. Thanh, E. Brück, O. Tegus, J. C. P. Klaasse, T. J. Gortenmulder, K. H. J. Buschow, J. Appl. Phys. **99**, 08Q107 (2006).
[21] Xue Chaofan, Yong Wang, Zhi Guo, Yanqing Wu, Xiangjun Zhen, Min Chen, Jiahua Chen, Song Xue, Zhongqi Peng, Qipeng Lu, and Renzhong Tai, Rev. Scien. Instr. **81**, 103502 (2010).